\documentstyle[sprocl,epsfig]{article}

\def\Journal#1#2#3#4{{#1} {\bf #2}, #3 (#4)}

\def\NCA{\em Nuovo Cimento}

\def\NPB{{\em Nucl. Phys.} B}

\def\PLB{{\em Phys. Lett.}  B}

\def\PRL{\em Phys. Rev. Lett.}

\def\PRD{{\em Phys. Rev.} D}

\def\ZPC{{\em Z. Phys.} C}

\def\PAN{\em Phys. Atom. Nucl.}

\def\al{\alpha}

\def\be{\begin{equation}}

\def\ee{\end{equation}}

\def\bea{\begin{eqnarray}}

\def\eea{\end{eqnarray}}

\begin{document}

\title{WHICH SCALAR MESON IS THE GLUE-STATE ?}

\author{M. BOGLIONE}

\address{
Dipartimento di Fisica Teorica, Universita' di Torino,\\ 
I.N.F.N, Sezione di Torino,
Via P.Giuria 1, I-10125 Torino, Italy \\
and\\
Centre for Particle Theory, University of Durham, \\ 
Durham DH1 3LE, U.K.}




\maketitle\abstracts{
Preliminary results of a work in collaboration with M.R. Pennington are 
presented.
Extending a scheme introduced by Tornqvist, we 
investigate a dynamical model in which the spectrum of scalar mesons can
be derived, with the aim of locating the lightest glue-state. Adding hadronic
interaction contributions to the bare propagator, to `dress' the bare 
quark-model $q \overline q$ states, we are able to write the amplitudes 
and the phase shifts in the approximation in which scalar resonances decay 
only into two pseudoscalar channels. The fit of these quantities to 
experimental data gives a satisfactory understanding of how
hadronic interactions modify the underlying `bare' 
spectrum. In particular, we examine the case in which a glue-state is 
introduced into the model.  
}

\section{Introduction}

Gluons carry colour charge which means they interact together. Consequently
it is possible for them to cluster 
and form objects which are colourless overall. These would be just like 
conventional hadrons, but with constituents that are massless gauge bosons.
These are known as {\em glueballs}. \\
Interest in glueballs has increased since new candidates for gluonic 
states have emerged 
from experimental results \cite{crb}$^{\!-\,}$\cite{mark} especially 
in the $1.5 - 2.0$ GeV energy region.
Moreover, Lattice QCD calculations (in quenched approximation) have 
recently suggested the presence 
of light scalar glue-states in this same energy region \cite{wein}.
From the experimental point of view, it is now clear that there are too
many confirmed scalar ($0^{++}$) 
mesons to form one $q \overline q$  meson nonet. The problem is
particularly pronounced 
for the I=0 sector, since at least four $f_0$'s now appear in 
the particle data listing. As a consequence, we can infer that
some of them have to be extra states.\\
The quark model gives a reasonably good description of the vector and
tensor meson spectra and properties, but 
its predictions for the scalar sector are very disappointing. 
To understand how and why scalars are so different from vectors and
tensors, we consider a simple 
model in which all bare meson states belong to ideally mixed quark multiplets. 
We call $n \overline n$ the 
nonstrange light state and suppose that substituting a strange quark for a 
light one increases the mass of the state by $\Delta m_s \simeq 100$ MeV, 
as illustrated in Fig.~1.\\

\begin{figure}[ht]
\vspace{-1.0cm}
\begin{center}
\mbox{~\epsfig{file=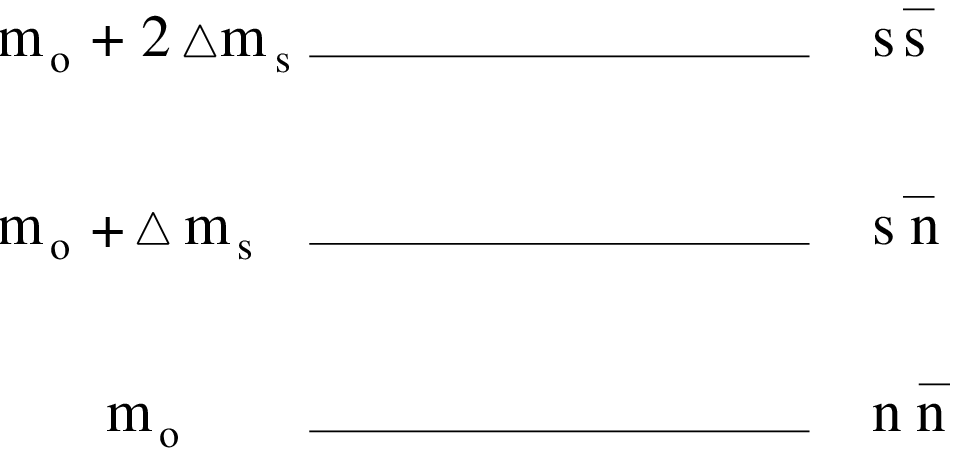,angle=-90,width=4.0cm}}
\vspace{.5cm}
\caption{
Spectrum of the bare $q \overline q$ states}
\end{center}
\end{figure}

\noindent
The bare propagator for each of these bound states will be of the form 
\cite{mike}
\be
P \,=\, \frac{1}{{\cal{M}} _0^2 - s } \; ,
\ee
with a pole on the real axis, corresponding to a non decaying state, 
for example
\[ |\phi \rangle _0 = |s \overline s \rangle \]  for the vector $I=0$
state, and 
\[|f_0 ' \, \rangle _0 = |s \overline s \rangle \] for the scalar  $I=0$ state.
If we now assume that the experimentally observed hadrons  are obtained from
 the bare states ($n \overline n$, $s \overline n$, $s \overline s$, ...)
by dressing them with hadronic interactions, the propagator becomes
\be 
P(s) \,=\, \frac{1}{{\cal{M}} ^2(s) - s -i \, {\cal{M}} (s) \, 
\Gamma (s)}  \; ,
\ee
and the pole moves into the complex $s$-plane. The corresponding state
can be decomposed as
\be 
|\phi \rangle = \sqrt{1-\epsilon ^2}|s \overline s \rangle + \epsilon _1 
|K \overline K \rangle 
 + \epsilon _2 |\rho \pi \rangle  + ...
\ee
where calculation would give \( \epsilon ^2 = \epsilon _1 ^2 + \epsilon _2
^2 + ... \ll 1 \).
The hadronic loop contributions allow the bare states ($s \overline s$ in
this example) to communicate with all hadronic channels permitted by quantum
numbers, and this enables the $\phi$ meson to decay.
A similar picture works for the tensors.\\
For scalars the situation is different because the dominant decays are just
into two pseudoscalars, the couplings  
are bigger and they couple strongly to more than one 
channel, creating overlapping and interfering resonance structures. Furthermore,
being S-waves, the opening at thresholds produces a more dramatic 
$s$-dependence in the propagator. As a consequence, the $f_0(980)$, 
for example, turns out to be 
predominantly a $|K \overline K \rangle$ state, and not an  
$|s \overline s \rangle$ one:
\be 
|f_0(980) \rangle \,=\, \sqrt{1-\epsilon ^2} \, |K \overline K \rangle + 
\epsilon _1 |s \overline s 
\rangle + ... \; \; .
\ee
This does not mean that the $f_0(980)$ is a $K \overline K$ molecule,
because 
the {\em seeds} of the model are conventional $q \overline q$ states and
the binding forces are not due to inter-hadron 
interactions alone. The crucial point here is the fact that hadronic 
interactions allow the 
$q \overline q$ bare states to communicate with all possible hadronic 
channels and these channels to communicate with each other, giving rise to a
{\em mixing} which means, for example, that the 
$f_0(980)$ spend most of its time in a $K \overline K$ state and not in
an $s \overline s$ one.\\

\section{A closer look at the model}

Let us now examine the model in more detail starting, for simplicity,
from the case in which just one resonance is produced.
If we define a vacuum polarization function $\Pi (s)$ which takes into
account all the possible two pseudoscalar 
loop contributions to the propagator $P(s)$, we can easily write
its imaginary part \cite{torn}:
\be
{\rm Im} \Pi (s) \,=\, - \sum _{i} G_i ^2 (s) \,=\,  
- \sum _{i} g_i ^2 \, \frac{k_i(s)}{\sqrt s} \, (s -s_{A,i}) \,  F_i ^2(s)
\, 
\theta (s-s_{th,i})
\ee
where the index $i$ runs over the pseudoscalar channels, the
$g_i$'s are the $SU(3)$ flavour couplings, the 
$k_i$'s are the c.m. momenta and $(s-s_{A,i})$ are the Adler zeros.
$F_i(s)$ are the form factors, which take into account the fact that the 
interaction is not 
pointlike but has a spatial extension. These are parametrized by
\be
F_i \,=\,\exp \; \left( \frac{-k_i ^2 (s)}{2\,k_0 ^2} \right) \; ,
\ee
where the momentum $k_0$ is inversely proportional to the range of the 
interaction. The real part of the vacuum polarization function can be found
from the dispersion relation 
\be
{\rm Re} \, \Pi (s) \,=\, \frac{1}{\pi} P \int _{s_{th,1}} ^\infty ds' \; 
\frac{{\rm Im} \Pi (s')}{s'-s}
\ee
No subtraction is needed, since the form factors decrease fast enough when
$|s| \to \infty$. 
At this point we are able to write the propagator as 
\be
P(s) \,=\, \frac{1}{m_0 ^2 + \Pi (s) - s } \; ,
\ee
and the contribution to the $i \to j$ amplitude as
\be\,\,
R_{ij} (s) \,=\, \frac{G_i(s)\,G_j(s)}{m_0 ^2 + \Pi (s) - s} \; . 
\label{R}
\ee
Notice that $R_{ij}(s)$, having its numerator and denominator related to
each other by the same $G_i$ couplings, is not the most general amplitude 
satisfying unitarity. This would look like:
\be
A_{ij} (s) \,=\, R_{ij} (s) \,  e ^{\,2 i \, \al (s)} + \sin \al (s) \;
e^{\,i \, \al (s)}
\label{A}
\ee
where $\alpha (s)$ is an unknown function of $s$ real along the right hand cut,
and the second term in Eq. (8) accounts for the background contribution.
To avoid an increasing number of parameters in the model, we will 
consider just the 
approximate amplitude $R_{ij}$ of Eq. (\ref {R}), having checked that a
constant value of $\alpha $ of about $15 ^{\circ}$  in Eq. (\ref {A}) would
typically work just as well.\\

\noindent
Fig. 2 shows the behaviour of the real and imaginary part of the running
complex mass function $m^2(s) = m_0^2 + \Pi (s)$, for the $I=1/2$, 
$K_0^*(1430)$, and the $I=1$, $a_0(980)$. Here the
parameters of the model (the mass of the bare $n\overline n$ state, the 
form factor cut-off $k_0$, an overall coupling $\gamma$ and the position of
the Adler zero) are determined by fitting the amplitudes and the phase
shifts to the experimental data from LASS \cite{lass}. 
As we can see, the opening
of each threshold gives an extra contribution (square root cusp) to the 
imaginary part of the vacuum polarization function. This is reflected in
the shape of its real part, which presents a very strong $s$-dependence at 
every threshold. The intersection of ${\rm Re} \, m^2(s)$ with the curve
$s$ represents the square of the Breit Wigner mass, and the value of 
${\rm Im} \, \Pi (s)$ 
at this point is related to the Breit Wigner width by 
\be
\Gamma _{BW} = \frac{- {\rm Im} \, \Pi (m_{BW})}{m_{BW}} \;\; .
\ee

\begin{figure}[ht]
\vspace{-1.0cm}
\hspace{-0.5cm}
\mbox{~\epsfig{file=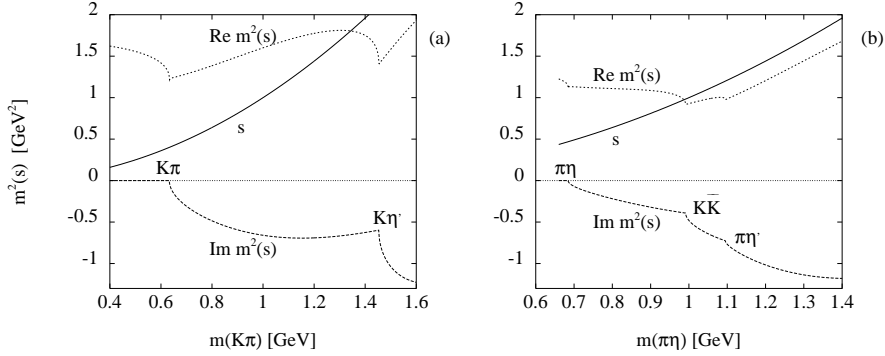,angle=-90,width=10.5cm}}
\vspace{-2.0cm}
\caption{\leftskip = 0.3cm 
Real and imaginary parts of the complex running mass functions for the
(a) $K_0^*(1430)$, and (b) $a_0(980)$, illustrating the effect of thresholds.}
\end{figure}

\noindent
Notice how the negative contribution of ${\rm Re} \, \Pi (s)$ shifts down the
actual mass of the resonance with respect to the value of the corresponding
bare mass. 
This effect is particularly pronounced for the $a_0(980)$ because the mass 
of the dressed bound state 
happens to coincide with the first of two thresholds, the $K \overline K$
and the $\pi \eta
'$ ones, which are very close to each other and have similar couplings: 
despite the bare mass $m_0$ being fixed at $1420$ MeV, the Breit Wigner
mass of the $a_0$ is found to be as low as $987$ MeV.\\

\newpage

\noindent
To analyze the more controversial $I=0$ scalar sector, we need to examine
the case in which more than one resonance is produced. Now, the polarization
function has to incorporate the features (couplings, masses,
form factors, ...) of each of the intermediate particles created in the process:
\be
{\rm Im} \Pi _{\al \beta} (s) \,=\, - \sum _{i} G_{\al,i}(s) \,
G_{i,\beta}(s) \; ,  
\ee
\be
{\rm Re} \, \Pi _{\al \beta} (s) \,=\, \frac{1}{\pi} P \int _{s_{th,1}} 
^\infty ds' \; 
\frac{{\rm Im} \Pi _{\al \beta} (s')}{s'-s} \, ,
\ee 
where the indices $\alpha$ and $\beta$ run over the $N$ different resonances. As
a consequence, the propagator 
\be
P _{\al \beta} (s) \,=\, \frac{1}{(m_0 ^2 - s)\delta _{\al \beta} + 
\Pi _{\al \beta} (s) } \; 
\ee
becomes an $N \times N$ complex matrix, and the couplings are
$N$-dimensional column vectors. 
The amplitudes are determined using a diagonalization procedure and can be 
written as
\be
R_{ij} (s) \,=\, \sum _{\al} \frac{G'_{\al,i}(s)\,G'_{j,\al}(s)}
{m_{\al, diag} ^2 (s) - s} 
\ee
where the diagonalized masses $m_{\al, diag} (s)$ are admixtures of all the
bare masses and the vacuum polarization matrix elements. They are, like the
new couplings $G \, '_{\al,i}(s)$, complex and $s$-dependent. \\
What is the interpretation of this more complicated picture ?
The physical observed hadrons are the states we obtain after
diagonalization, an admixture of the {\em seed } states of the
model. 
The mixing among them is embodied in the diagonalization procedure, 
which allow
all the channels to communicate with each other and to create new physical 
states with masses, couplings and widths different from the ones of the 
primitive states.\\

\noindent
We now want to consider explicitly the case in which not only the two
conventional $f_0$ and $f_0'$ are present in the $I=0$ sector, but also a
third state arises, thanks to the presence of a glue $gg$ seed.
Since, in principle, we do not know the mass of the bare glue-state,
$m_{gg}$ is assumed to be an extra parameter of the model.
To avoid a further increase in the number of parameters we will also assume 
that the $gg$ state cannot mix with the $q \overline q$ states at 
the bare level: all the mixing will occur via hadron interactions. 
The bare couplings for the glueball to the two pseudoscalar channels are
the ones of an $SU(3)$ flavour singlet. 
Fig.~3 presents the results we obtain if we choose a $gg$ bare mass 
of $1.8$ GeV and readjust the other parameters to fit the data 
from $\pi \pi$ scattering~\cite{ochs}$^{\!,\,}$\cite{cason}, in such a way 
that the agreement
with the LASS data on $K_0^*(1439)$ \cite{lass} remains  satisfactory. 
The two lower states have the same features that 
would be obtained if only two resonances were considered 
\cite {torn}: a very broad state with a mass of about 1 GeV and a narrow
one with a mass of about 1.2 GeV. The higher glue-state almost decouples 
from the 
lowest channels, but shows relatively strong couplings not only to the 
heaviest $\eta' \eta'$, but also to the $\eta \eta'$ channel, to which it had 
a zero underlying bare coupling. Again, the Breit Wigner 
mass is considerably lower then the input one: for these values of the
parameters our glueball mass is about 
$1.57$ GeV and its width is about $200$ MeV. Notice how the masses of all
the scalar $I=0$ mesons are related to
the positions of the two pseudoscalar thresholds.  
The amplitude presents two dips: the former corresponds 
to the presence of a very narrow resonance sitting on top of a very broad 
one, the second is related to the third heavier resonance, in agreement with 
the experimental data from Crystal Ball \cite{crb}, GAMS \cite{kond}, and
the analysis from  Bugg et al. \cite{bugg} and Anisovich et al.
\cite{anis}\\

\newpage

\begin{figure}[ht]
\vspace{-1.0cm}
\hspace{-0.5cm}
\mbox{~\epsfig{file=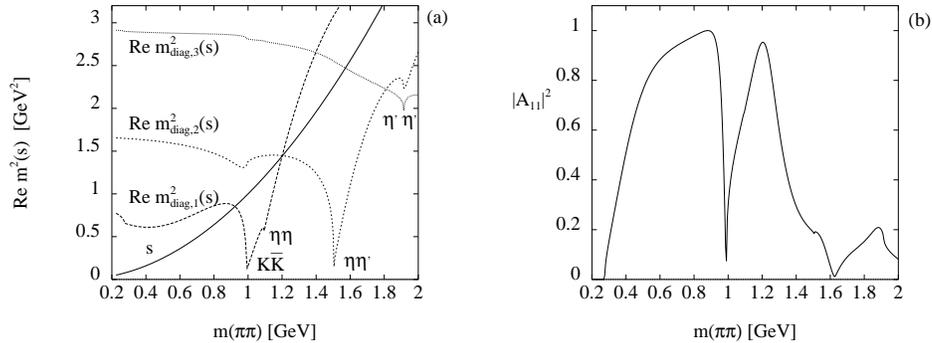,angle=-90,width=10.5cm}}
\vspace{-2.0cm}
\caption{\leftskip = 0.4cm 
(a) Real parts of the diagonalized masses of 
the three $I=0$ scalar resonances created in $\pi \pi$ scattering. 
(b) Modulus of the amplitude corresponding to the same process}
\end{figure}

\section{Conclusions}

In this talk I have presented some preliminary results of a work performed in
collaboration with M.R.Pennington, which is still in 
progress. In particular, a proper fit of the more recent 
experimental data up to about $1.8$ GeV \cite{lesn}$^{\!,\,}$\cite{kond} 
has still to be completed.\\ 
The model we are using requires approximations, to allow the calculation to
be performed. The amplitudes are assumed to be pole dominated
and so the unitarization chosen is not the most general one. Furthermore, only
decays into two pseudoscalars are considered, whereas other thresholds 
like multipion, vector-vector or axial-pseudoscalar ones are  
neglected. \\
Despite the assumptions, this scheme has many great advantages: 
first of all it gives a 
{\em dynamical} description of the particularly complex mechanism which leads 
to the creation of scalar resonances, naturally taking into account the
mixing amongst different states, and explaining why the 
scalars differ from vector and tensor mesons. Moreover, it is simple
and with very few parameters.\\ 
Consequently, we believe this model is not a  
machine from which one can blindly extract numbers from a fitting program, 
but an efficient schematic way to approach the 
{\em dynamics} of the non-perturbative hadronic world: a world in which so
much remains to be understood.\\

\section*{Acknowledgments}

I am very grateful to M.R. Pennington for guiding and supervising me with
great expertise and infinite patience. This contribution would not have been
possible without his help and support. 
I thank M. Anselmino and E. Predazzi for having invited me to Turin for this
interesting and entertaining meeting, and D. Lichtenberg for useful
discussions and delightful company. 
Finally, I acknowledge I.N.F.N, Sezione di Torino, for travel support.

\end{document}